\documentclass[sigconf]{acmart}

\AtBeginDocument{%
  \providecommand\BibTeX{{%
    \normalfont B\kern-0.5em{\scshape i\kern-0.25em b}\kern-0.8em\TeX}}}

\copyrightyear{2024}
\acmYear{2024}
\setcopyright{acmlicensed}\acmConference[KDD '24]{Proceedings of the 30th ACM SIGKDD Conference on Knowledge Discovery and Data Mining}{August 25--29, 2024}{Barcelona, Spain}
\acmBooktitle{Proceedings of the 30th ACM SIGKDD Conference on Knowledge Discovery and Data Mining (KDD '24), August 25--29, 2024, Barcelona, Spain}
\acmDOI{10.1145/3637528.3671840}
\acmISBN{979-8-4007-0490-1/24/08}

\acmSubmissionID{950}

\usepackage{graphicx}
\usepackage{amsfonts}
\usepackage{amsmath}
\usepackage{bm}
\usepackage{booktabs}
\usepackage{multirow}
\usepackage[flushleft]{threeparttable}
\usepackage{balance}
\usepackage{caption}
\usepackage{hyperref}

\definecolor{mygray}{gray}{0.9}

\newcommand{\e}{\bm{e}}
\newcommand{\A}{\mathbf{A}}

\begin{document}

\title{Unifying Graph Convolution and Contrastive Learning in Collaborative Filtering}

\author{Yihong Wu}
\affiliation{%
  \institution{Université de Montréal}
  \city{Montréal}
  \country{Canada}}
\email{yihong.wu@umontreal.ca}

\author{Le Zhang}
\affiliation{%
  \institution{Mila - Quebec AI Institute}
  \city{Montréal}
  \country{Canada}}
\email{le.zhang@mila.quebec}

\author{Fengran Mo}
\affiliation{%
  \institution{Université de Montréal}
  \city{Montréal}
  \country{Canada}}
\email{fengran.mo@umontreal.ca}

\author{Tianyu Zhu}
\authornote{This work was done at Université de Montréal.}
\affiliation{%
 \department{MIIT Key Laboratory of Data Intelligence and Management}
 \institution{Beihang University}
 \city{Beijing}
 \country{China}}
\email{ztybuaa@126.com}

\author{Weizhi Ma}
\affiliation{%
  \institution{Institute for AI Industry Research (AIR), Tsinghua University}
  \city{Beijing}
  \country{China}}
\email{mawz@tsinghua.edu.cn}

\author{Jian-Yun Nie}
\affiliation{%
  \institution{Université de Montréal}
  \city{Montréal}
  \country{Canada}}
\email{nie@iro.umontreal.ca}

\renewcommand{\shortauthors}{Yihong Wu et al.}

\begin{abstract}
Graph-based models and contrastive learning have emerged as prominent methods in Collaborative Filtering (CF).
While many existing models in CF incorporate these methods in their design, there seems to be a limited depth of analysis regarding the foundational principles behind them.
This paper bridges graph convolution, a pivotal element of graph-based models, with contrastive learning through a theoretical framework. By examining the learning dynamics and equilibrium of the contrastive loss, we offer a fresh lens to understand contrastive learning via graph theory, emphasizing its capability to capture high-order connectivity.
Building on this analysis, we further show that the graph convolutional layers often used in graph-based models are not essential for high-order connectivity modeling and might contribute to the risk of oversmoothing.
Stemming from our findings, we introduce Simple Contrastive Collaborative Filtering (SCCF), a simple and effective algorithm based on a naive embedding model and a modified contrastive loss. The efficacy of the algorithm is demonstrated through extensive experiments across four public datasets.
The experiment code is available at \url{https://github.com/wu1hong/SCCF}.
\end{abstract}




\maketitle

\section{Introduction}
Recently, contrastive learning has become the state-of-the-art method for self-supervised learning, marking significant achievements across computer vision \cite{wu2018unsupervised, chen2020simple, he2020momentum, grill2020bootstrap}, natural language processing \cite{gao2021simcse}, and multi-modality \cite{radford2021learning}.
Given its success in various domains, there is burgeoning interest in harnessing contrastive learning within Collaborative Filtering (CF) \cite{wang2022towards,lee2021bootstrapping, zhou2021contrastive, zhang2022incorporating}.
\citet{wu2022effectiveness} explored the effectiveness of the Sampled SoftMax loss function within CF. \citet{chen2023adap} examined the impact of the temperature parameter on a model's performance in contrastive learning. Both \citet{zhou2021contrastive} and \citet{zhang2022incorporating} highlighted the ability of contrastive loss to address bias in recommendation applications. \citet{zhou2023contrastive} employed contrastive learning to address cold-start problems in recommendation.
Yet, none of these studies address the fundamental question: \textit{How does contrastive learning operate within collaborative filtering}?
While studies such as \citet{wang2020understanding} and \citet{wang2022towards} have elucidated that contrastive learning serves to align the embeddings of interacted user-item pairs and to uniformly distribute embeddings across a hypersphere, the mechanism and equilibrium of these forces remain veiled.

Beyond contrastive learning, graph-based methods constitute another burgeoning research avenue in CF. Drawing inspiration from the success of Graph Convolutional Networks (GCN) \cite{kipf2016semi}, CF researchers have discovered that integrating graph convolutional layers with basic learnable embeddings significantly enhances recommendation quality \cite{wang2019neural, he2020lightgcn}. While instances exist where researchers simultaneously employ contrastive learning and graph-based approaches \cite{wang2022towards, mao2021ultragcn}, these domains are often treated as distinct and separate, resulting in a lack of comprehensive exploration. Motivated by the recent breakthrough on a theoretical link between contrastive learning and graph theory \cite{haochen2021provable, tan2023contrastive, wang2022chaos, wang2023message}, this paper aims to bridge the existing gap in CF research and address the fundamental questions of contrastive learning through the framework of graph theory.

Specifically, we establish the equivalence between contrastive learning and graph convolution within the context of collaborative filtering. We demonstrate that the learning dynamics (embedding update) of the contrastive loss constitute a combination of two distinct graph convolution processes.
One graph convolution operation, represented by positive samples, enhances the smoothness of embeddings across the graph. Conversely, another convolution operation, characterized by negative samples, reduces this smoothness, thereby preventing the collapse of embeddings.
These two operations serve as opposing forces within the system, dynamically attracting and dispersing embeddings, respectively.
Furthermore, achieving equilibrium (a state of balance) within these two forces necessitates the model's estimations to align with the empirical distribution.

A significant advantage of this equivalence is the ability of contrastive learning to effectively model High-Order Connectivity (HOC) \cite{wang2019neural}, a capability traditionally pursued through the incorporation of graph convolutional layers in most existing graph-based methods \cite{wang2019neural, he2020lightgcn, wu2021self}. However, our findings challenge the necessity of graph convolutional layers for HOC modeling, indicating that an increase in the number of these layers may lead to oversmoothing. In response to the limitations posed by convolutional layers and leveraging the potential of contrastive loss for HOC modeling, we propose the Simple Contrastive Collaborative Filtering (SCCF) approach.
This approach comprises a naive embedding model and a tailored contrastive loss function.
Without any convolutional layer, our minimalist model attains performance that matches or even surpasses state-of-the-art methods in collaborative filtering.

Our contributions are summarized as follows: (1) We present a theoretical framework that integrates contrastive learning with graph convolution. Within this framework, the contrastive loss is decomposed into two distinct convolution processes: one that attracts embeddings closer together and another that disperses them. (2) We highlight the capacity of contrastive loss to model High-Order Connectivity (HOC) and discuss the necessity of graph convolutional layers. (3) We propose the Simple Contrastive Collaborative Filtering (SCCF) method, consisting only of a naive embedding model and a contrastive objective. Extensive experiments confirm its efficacy and its ability to model HOC.

\section{Preliminaries}
\subsection{Task Definition}
Consider a classic CF setting with implicit feedback for Top-K recommendation, where only binary values indicating the interactions between users and items are available.
Our goal is to model a similarity function $s(u,i)$ reflecting the degree of interest user $u$ has in item $i$.
In a Top-K recommendation scenario, the recommender presents a user with K items, selected based on their similarity to the user’s demonstrated interests.

\subsection{Notation}
We define some necessary notations for this paper.
Let $\mathcal{U}$ represent the set of users and $\mathcal{I}$ denote the set of items.
Let dataset $\mathcal{D}= \{(u_1, i_1), (u_2, i_2), ..., (u_n, i_n)\}$ denote the set of observed interactions between users and items, $\mathcal{D}_U = \{u_1, u_2, \dots, u_n\}$ denote the multiset of users observed in the dataset, $\mathcal{D}_I = \{i_1, i_2, \dots, i_n\}$ denote the multiset of items, $|\mathcal{D}|=n$  be the number of interactions.
The interactions between users and items are represented by a matrix $\mathbf{R} \in \mathbb{R}^{|\mathcal{U}| \times |\mathcal{I}|}$ as follows:
\begin{equation}
    \mathbf{R}_{ui}=\begin{cases}1,\;(u,i)\in \mathcal{D};\\0,\;(u,i)\notin \mathcal{D}.\\ \end{cases}
\end{equation}
The interactions of users and items can be represented by a bipartite graph \cite{zha2001bipartite}, whose adjacency matrix $\A \in \mathbb{R}^{(|\mathcal{U}|+|\mathcal{I}|)\times (|\mathcal{U}|+|\mathcal{I}|)}$ is defined as
\begin{equation}
\mathbf{A} = 
    \begin{pmatrix}
\mathbf{0} & \mathbf{R}\\
\mathbf{R}^\top & 0
\end{pmatrix}.
\end{equation}
Let $\mathbf{D}=diag(d_1, d_2, \dots, d_n)$ denote the degree matrix where $d_i$ is the degree of node $i$.
Let $\mathbf{L}= \mathbf{D}-\mathbf{A}$ denotes the Laplacian matrix.
We denote the embedding of user $u$ and item $i$ as $\e_u$ and $\e_i$, respectively, regardless of the encoder and use the inner product function to decode the similarity from embeddings, i.e., $s(u,i)=\e_u^\top\e_i$.

\subsection{Graph Convolution}
\label{intro_graph_convolution}
Consider a graph $\mathcal{G}=(\mathcal{V}, \mathcal{E})$ with $n$ nodes, where $\mathcal{V}$ is the set of vertices and $\mathcal{E}$ is the set of edges, and its adjacency matrix $\mathbf{A}\in \mathbb{R}^{n \times n}$.
A \textbf{graph signal} is a function $\bm{x}: \mathcal{V}\rightarrow \mathbb{R}$ mapping a node to a real value.
Intuitively, a smooth signal should share similar values across connected nodes.
Consequently, we use the \textit{normalized graph quadratic form} \cite{ramakrishna2020user} to measure the smoothness of a graph signal defined as $S(\boldsymbol{x})=\bm{x}^{\top} \mathbf{L} \bm{x} / \|\bm{x}\|_2 = \sum_{i, j} A_{i j}\left(x_i-x_j\right)^2 / \|\bm{x}\|_2.$
A low value $S(\bm{x})$ indicates a smooth signal $\bm{x}$. 
Since the Laplacian matrix $\mathbf{L}$ is real and symmetric, the eigendecomposition yields $\mathbf{L} = \mathbf{U}\Lambda\mathbf{U}^\top,$
where $\Lambda=diag(\lambda_1, \lambda_2, \dots, \lambda_n)$ is a diagonal matrix whose entries are eigenvalues with $\lambda_1 \leq \lambda_2 \leq \dots \leq \lambda_n$, $\mathbf{U}=[ \bm{u}_1, \bm{u}_2, \dots, \bm{u}_n]$ is a matrix of eigenvectors, $\bm{u}_i \in \mathbb{R}^n$ is the unitary eigenvector corresponding to eigenvalue $\lambda_i$.
Observing that $ S(\bm{u}_i) = \lambda_i$, it can be inferred that eigenvectors associated with smaller eigenvalues tend to be smoother.
This observation motivates us to define the \textbf{graph frequencies} as the Laplacian eigenvalues and the Graph Fourier Transformation (GFT) based on the eigenvectors $\mathbf{U}$ \cite{sandryhaila2013discrete}.
The GFT of signal $\bm{x}$ is represented as $\hat{\bm{x}} =\mathbf{U}^\top\bm{x}$.
This transformation maps any signal $\bm{x}$ into the graph frequency space, in which
$\hat{\bm{x}}$ serves as the new coordinate of $\bm{x}$ and reflects the importance of frequencies.
To end this section, we give the definition of \textbf{graph filter} and \textbf{graph convolution}.
\begin{definition}[Graph Filter]
    Given a graph and its corresponding Laplacian matrix, eigenvalues, and eigenvectors, the graph filter $\mathcal{H}(\bm{L})$ is defined as the following
    \begin{equation}
        \mathcal{H}(\bm{L}) = \sum_{i=0}^\infty h_i \mathbf{L}^i,
    \end{equation}
    where $h_i$ is the coefficient for $\mathbf{L}^i$.
\end{definition}
\begin{definition}[Graph Convolution]
Given a graph signal $\bm{x}$ and a graph filter $\mathcal{H}(\bm{L})$, graph convolution represents the process of applying a graph signal to a graph filter, i.e., $\mathcal{H}(\bm{L})\bm{x}$.
\end{definition}

\section{Unifying Graph Convolution and Contrastive Learning}
In this section, we begin by defining the contrastive loss function for analysis. Subsequently, we examine the learning dynamics of the contrastive loss and its relationship to graph convolution. Following this, we identify the conditions necessary to achieve equilibrium within the contrastive loss framework. The section concludes with a discussion on alternative forms of loss functions.
\label{unifying}

\subsection{Definition of Contrastive Loss}
In the context of contrastive learning, defining positive and negative samples with respect to a given anchor is crucial. In the field of Computer Vision, positive samples are typically generated by augmenting the same image, whereas augmentations from different images are considered negative pairs \cite{chen2020simple, he2020momentum}. Conversely, in CF, positive samples for a user are identified as items with which the user has interacted, while items with no interaction history are deemed negative. 
Perhaps the most widely adopted contrastive loss function in the domain of collaborative filtering is the Sampled SoftMax (SSM) function \cite{wu2022effectiveness, chen2023adap, zhang2022incorporating}:
\begin{equation}
\label{ssm}
l_{\text{SSM}} = -\frac{1}{|\mathcal{D}|} \sum_{(u, i) \in \mathcal{D}} \log \frac{\exp (\e_u^\top \e_i)}{\sum_{j \in \mathcal{D}_i} \exp (\e_u^\top \e_j)}.
\end{equation}
Assuming the joint probability of a user \(u\) showing interest in an item \(i\), denoted as \(p(u, i)\), is proportional to $\exp (\e_u^\top \e_i)$, the expression $\exp (\e_u^\top \e_i) / \sum_{j \in \mathcal{D}_i} \exp (\e_u^\top \e_j)$ can be interpreted as the conditional probability of user \(u\) interacting with item \(i\), given the user, \(p(u,i|u)\). This formulation implies that the SSM function is designed to maximize the log-likelihood of this conditional probability for a given user. To enable a more straightforward analysis, we propose an alternative loss function that directly maximizes the log-likelihood of the joint probability from the observed data:
\begin{align}
    \label{infonce}
    \begin{split}
    l &= -\frac{1}{|\mathcal{D}|} \sum_{(u,i)\in \mathcal{D}} \log \frac{\exp \left(\bm{e}_u^\top \bm{e}_i\right)}{\sum_{(x, y)\in \mathcal{D}_U \times 
 \mathcal{D}_I} \exp \left(\bm{e}_x^\top \bm{e}_y\right)} \\
 &= -\frac{1}{|\mathcal{D}|} \sum_{(u,i)\in \mathcal{D}} \log \frac{\exp \left(\bm{e}_u^\top \bm{e}_i\right)}{\sum_{(x, y)\in \mathcal{U} \times 
 \mathcal{I}} d_xd_y\exp \left(\bm{e}_x^\top \bm{e}_y\right)}
    \end{split}
\end{align}
In Equation (\ref{infonce}), for a positive pair $(u,i)$, we consider all other possible combinations between $\mathcal{D}_U$ and $\mathcal{D}_I$ as negative pairs.
This negative sampling strategy is equivalent to the batch negative sampling trick in the case of unlimited batch size.
Moreover, the denominator $\sum_{(x, y)\in \mathcal{D}_U \times 
 \mathcal{D}_I} \exp \left(\bm{e}_x^\top \bm{e}_y\right)$ can be streamlined by focusing on unique user-item pairs and their co-occurrence frequencies. This leads to a simplified expression $\sum_{(u,i)\in \mathcal{\mathcal{U}\times\mathcal{I}}} d_u d_i\exp \left(\bm{e}_u^\top \bm{e}_i\right)$, which accounts for the redundancy of pairs in the original formulation.

\subsection{Learning Dynamics of Contrastive Loss}
\label{learning_dynamic}
In this section, we will derive the learning dynamic for the contrastive loss.
According to a recent analysis on contrastive learning \cite{wang2020understanding}, a contrastive loss encompasses two principal components: alignment and uniformity. The alignment component is designed to minimize the distance between positive user-item pairs within the embedding space, thereby enhancing their similarity. Conversely, the uniformity component aims to increase the distance among negative pairs, thereby preventing the embeddings from converging to a singular point.
Furthermore, this decomposition simplifies our discussion since the analysis of both the alignment and uniformity components are congruent.
We decompose Equation (\ref{infonce}) as follows:
\begin{align}
    l &= l_{\text{align}} + l_{\text{uniform}}, \\
    l_{\text{align}} &= -\frac{1}{|\mathcal{D}|} \sum_{(u,i)\in \mathcal{D}}\bm{e}_u^\top \bm{e}_i ,\label{eq_align} \\
    l_{\text{uniform}} &= \log \sum_{(u,i)\in \mathcal{\mathcal{U}\times\mathcal{I}}} d_u d_i\exp \left(\bm{e}_u^\top \bm{e}_i\right). \label{eq_uniform}
\end{align}
We commence by examining the learning dynamics associated with the alignment loss function, as described by Equation (\ref{eq_align}). The derivative of this function can be articulated as:
\begin{equation}
\begin{aligned}
    \frac{\partial l_{\text{align}}}{\partial \bm{e}_u} = -\frac{1}{|\mathcal{D}|}\sum_{i\in \mathcal{N}(u)} \bm{e}_i,\;\;
    \frac{\partial l_{\text{align}}}{\partial \bm{e}_i} = -\frac{1}{|\mathcal{D}|}\sum_{u\in \mathcal{N}(i)} \bm{e}_u,
\end{aligned} 
\end{equation}
where $\mathcal{N}(u)$ denotes the set of neighbors of user $u$ and $\mathcal{N}(i)$ for item $i$.
The gradient descent update of embeddings at step $t+1$ is
\begin{equation}
\label{update_rule}
\begin{aligned}
    \bm{e}_u(t+1) = \bm{e}_u(t) - \gamma \frac{\partial l}{\partial \bm{e}_u} = \bm{e}_u(t) + \frac{\gamma}{|\mathcal{D}|}\sum_{i\in \mathcal{N}(u)} \bm{e}_i(t),\\
    \bm{e}_i(t+1) = \bm{e}_i(t) - \gamma \frac{\partial l}{\partial \bm{e}_i} = \bm{e}_i(t) + \frac{\gamma}{|\mathcal{D}|}\sum_{u\in \mathcal{N}(i)} \bm{e}_u(t),
\end{aligned}
\end{equation}
where $\gamma$ is the learning rate.
Equation (\ref{update_rule}) delineates a message-passing mechanism, illustrating how embeddings are propagated across edges in the interaction graph.
For a more compact representation, we rewrite Equation (\ref{update_rule}) in a matrix form as
\begin{equation}
\label{align_update}
    \mathbf{E}(t+1) = \mathbf{E}(t) + \frac{\gamma}{|\mathcal{D}|} \mathbf{A}\mathbf{E}(t) = \left(\mathbf{I} + \frac{\gamma}{|\mathcal{D}|} \mathbf{A} \right)\mathbf{E}(t),
\end{equation}
where $\mathbf{E}(t) \in \mathbb{R}^{(|\mathcal{U}|+|\mathcal{I}|)\times d}$ is the concatenation of user and item embeddings at step $t$.
Similarly, we delve into the learning dynamics associated with the uniformity objective, as represented by Equation (\ref{eq_uniform}).
Let $Z = \sum_{(x,y)\in \mathcal{\mathcal{U}\times\mathcal{I}}} d_x d_y\exp \left(\bm{e}_x^\top \bm{e}_y\right)$.
The derivatives corresponding to this uniformity are expressed as:
\begin{equation}
\label{eq_12}
\begin{split}
    \frac{\partial l_{\text{uniform}}}{\partial \bm{e}_u} &= \sum_{i \in \mathcal{I}}\frac{d_u d_i\exp\left(\e_u^\top\e_i\right)}{Z} \e_i, \\
    \frac{\partial l_{\text{uniform}}}{\partial \bm{e}_i} &= \sum_{u \in \mathcal{U}}\frac{d_u d_i\exp\left(\e_u^\top\e_i\right)}{Z} \e_u.
\end{split}
\end{equation}
The embedding update for the uniformity in a matrix form is
\begin{equation}
\label{uniform_update}
    \mathbf{E}(t+1) = \mathbf{E}(t) - \gamma\mathbf{A}'(t)\mathbf{E}(t) = \left(\mathbf{I} - \gamma \mathbf{A}'(t)\right)\mathbf{E}(t),
\end{equation}
where
\begin{equation}
\label{a_prime}
    \mathbf{A}'_{ij}\left(t\right) =
        \frac{d_i d_j\exp\left(\bm{e}_i^\top(t) \bm{e}_j(t)\right)}{\sum_{(x,y)\in \mathcal{U}\times\mathcal{I}} d_x d_y\exp \left(\bm{e}_x^\top(t) \bm{e}_y(t)\right)} 
\end{equation}
if $(i,j)$ is a user-item or item-user pair; or $\mathbf{A}'_{ij}\left(t\right) = 0$ if $(i, j)$ is a user-user or item-item pair.
Equation (\ref{eq_12}) elucidates another message-passing mechanism across the complete graph weighted by their respective degrees and embeddings. Now we combine Equation (\ref{align_update}) and Equation (\ref{uniform_update}) to get the embedding update for Equation (\ref{infonce}):
\begin{equation}
\label{core}
    \mathbf{E}(t+1) = \mathbf{E}(t) + \gamma\mathbf{A}''(t)\mathbf{E}(t) = \left(\mathbf{I} + \gamma \mathbf{A}''(t)\right)\mathbf{E}(t),
\end{equation}
where
\begin{equation}
\label{eq_state}
    \mathbf{A}''(t) = \mathbf{A}/|\mathcal{D}| - \mathbf{A}'(t).
\end{equation}
Equation (\ref{core}) represents the whole learning dynamics for the contrastive loss defined in Equation (\ref{infonce}).

\subsection{Equivalence of Contrastive Loss and Graph Convolution}
In the preceding section, we derived the embedding update formula associated with the contrastive loss, Equation (\ref{infonce}). This section aims to demonstrate that the embedding update process is functionally equivalent to graph convolutions.
To begin with, we introduce some important propositions.
\begin{proposition}
\label{prop1}
    Given a graph and its corresponding Laplacian matrix $\mathbf{L}$, eigenvalues of $\mathbf{L}$ such that  $\lambda_1 \leq \lambda_2 \leq \dots \leq \lambda_n$, $ 0< \gamma < 1/\lambda_n$, graph filter $\mathbf{I} - \gamma \mathbf{L}$ is a low-pass filter and graph filter $\mathbf{I}+\gamma\mathbf{L}$ is a high-pass filter.
\end{proposition}
\begin{proposition}
\label{prop2}
    For any graph signal $\bm{x}$, low-pass filter $\mathcal{H}_L$, high-pass filter $\mathcal{H}_H$, graph convolution with low-pass filter $\mathcal{H}_L\bm{x}$ increases signal's smoothness on the graph, i.e., $S(\mathcal{H}_L\bm{x}) \leq S(\bm{x})$. Graph convolution with high-pass filter $\mathcal{H}_H$ decreases signal's smoothness, i.e., $S(\mathcal{H}_H\bm{x}) \geq S(\bm{x})$.
\end{proposition}
All the proofs can be found in \cite{isufi2024graph}.
By definition, a low-pass filter retains the low-frequency components of a graph signal while suppressing the high-frequency ones. As previously discussed in section \ref{intro_graph_convolution}, low-frequency components correspond to eigenvectors associated with smaller eigenvalues, thereby leading to smoother signals. Recall that we use the graph quadratic form $S(\bm{x})$ to measure the smoothness of graph signal $\bm{x}$. Applying a low-pass filter to a graph signal consequently results in a smoother outcome, with the converse holding for high-pass filters.
Interested readers are encouraged to consult \cite{isufi2024graph, ramakrishna2020user} for a more comprehensive understanding of graph filters. Having established these foundational propositions, we arrive at a theorem that articulates the equivalence between contrastive loss and graph convolution.
\begin{theorem}
\label{theorem1}
For a small enough learning rate $\gamma$, graph filter $\mathbf{I} + \gamma \mathbf{A} / |\mathcal{D}| $ increases signal's smoothness on the user-item interaction graph; graph filter $\mathbf{I} - \gamma \mathbf{A}'(t)$ decreases signal's smoothness on the affinity graph.
\end{theorem}
The proof can be found in Appendix \ref{proof_th1}.
With Theorem \ref{theorem1}, we deduce that the alignment loss, Equation (\ref{align_update}), functionally acts as a graph convolution to enhance the smoothness of embeddings on the user-item interaction graph. Conversely, the uniformity loss, Equation (\ref{uniform_update}), operates as a graph convolution to reduce the smoothness of embeddings on the affinity graph.

\subsection{Equilibrium of Contrastive Learning}
\label{equilibrium}
We demonstrate that the equilibrium in contrastive learning—the convergence of the learning process—necessitates the alignment of the model's estimation with the empirical distribution derived from the data. When the system stabilizes into this equilibrium, it adheres to the condition:
    $\mathbf{E}(\infty) = \mathbf{E}(\infty) + \gamma\mathbf{A}''\mathbf{E}(\infty)$.
From this, it follows that: $\mathbf{A}''\mathbf{E}(\infty) = \mathbf{0}$.
Since $\mathbf{E}$ cannot be a zero matrix $\mathbf{0}$, it follows that either $\mathbf{A}''$ is $\mathbf{0}$ or $\mathbf{E}$ is a null solution for $\mathbf{A}''$.
Summarizing the above reasoning, we have the following theorem:
\begin{theorem}
\label{theorem2}
    The contrastive loss reaches its equilibrium if and only if $\mathbf{A}' = \mathbf{A}/|\mathcal{D}|$
\end{theorem}
The proof can be found in Appendix \ref{proof_th2}.
Observe that $\mathbf{A}/|\mathcal{D}|$ can be considered as the empirical distribution of interactions between users and items denoted as $P_e(u,i)$. Formally, 
\begin{equation}
    \frac{\mathbf{A}_{ui}}{|\mathcal{D}|} = P_e(u, i) = 
    \begin{cases}
        1/|\mathcal{D}|, &(u,i) \in \mathcal{D};\\
        0,  &(u,i) \notin \mathcal{D}.\\
    \end{cases}
\end{equation}
This empirical distribution assigns weights uniformly across every observed user-item pairing, devoid of any prior assumptions.
On the other hand, $\mathbf{A}'(t)$ can be considered as the Boltzmann distribution reflecting the model's estimation:
\begin{equation}
    \mathbf{A}_{ui}'(t) = P_B(u, i) = \frac{d_u d_i\exp \left(\bm{e}_u^\top \bm{e}_i \right)}{Z},
\end{equation}
A Boltzmann distribution \cite{ackley1985learning} is a probability distribution estimated by embeddings through an energy function $\exp (\bm{e}_i^\top \bm{e}_j)$.
Theorem \ref{theorem2} articulates that achieving equilibrium necessitates the alignment of the model's estimated probability with the empirical probability.
Moreover, by solving Equation $\mathbf{A}' = \mathbf{A}/|\mathcal{D}|$, we obtain an expression for the similarity between the user and item
$\bm{e}_u^\top\bm{e}_i = \log \frac{\mathbf{A}_{ui}}{|\mathcal{D}|d_u d_i} + \log Z$.
This relationship can also be represented in matrix form:
\begin{equation}
\label{eq_mf_contrastive}
    \mathbf{E}\mathbf{E}^\top = \log{ \mathbf{D}^{-1}\mathbf{A}\mathbf{D}^{-1}} - \log{|\mathcal{D}|} + \log{Z}.
\end{equation}
Equation (\ref{eq_mf_contrastive}) reveals that optimization under the contrastive loss is equivalent to performing an implicit matrix factorization \cite{qiu2018network, levy2014neural}. This equivalence provides an alternative perspective to assess the quality of embeddings.

\subsection{Discussion}
In summary, we describe the learning dynamics of the contrastive loss and identify its equivalence with graph convolutions.
Moreover, the alignment loss corresponds to the graph convolution smoothing embeddings on the user-item interaction graph while the uniformity loss corresponds to the graph convolution dispersing embeddings on the affinity graph.
Lastly, we demonstrate that the equilibrium of contrastive loss requires the model's estimation to match the empirical distribution and that the optimization via the contrastive loss is an implicit matrix factorization.

Furthermore, our analysis paradigm (learning dynamics and equilibrium) is not limited to Equation (\ref{infonce}).
It could also be applied to the SSM function (Equation (\ref{ssm})), the Mean Squared Error (MSE) loss (Matrix Factorization \cite{koren2009matrix}), the Bayesian Personalized Ranking (BPR) loss \cite{rendle2012bpr}, or even the recently proposed DirecAU loss \cite{wang2022towards} which adopted negative samples as user-user and item-item pairs.

\section{High-Order Connectivity Modeling}
High-Order Connectivity (HOC) \cite{wang2019neural} (or High-Order Proximity \cite{yang2018hop}) modeling is a desired property for CF methods and has been the main motivation for graph-based CF methods.
The HOC modeling requires the similarity between node embeddings should reflect their closeness on the graph.
Traditional methods such as Matrix Factorization (MF) \cite{koren2009matrix} and BPR \cite{rendle2012bpr} are often considered inadequate for modeling HOC. This perceived limitation stems from these methods' primary focus on distinguishing observed items from unobserved ones, while largely neglecting to explicitly model the latent relationships among unobserved entities \cite{yang2018hop, wang2019neural}.
To mitigate this limitation, earlier graph-based collaborative filtering approaches have employed random walks to derive HOC scores. These scores are then directly utilized to inform recommendations \cite{paudel2016updatable, tong2006fast}.
HOR-rec \cite{yang2018hop} enhances the BPR loss function by integrating HOC through weighted coefficients, alongside an expanded set of positive and negative pairs derived from random walks. Subsequently, NGCF \cite{wang2019neural} and subsequent graph-based methods \cite{he2020lightgcn, wu2021self} draw upon the Graph Convolutional Network (GCN) framework \cite{kipf2016semi}, employing multiple convolutional layers to facilitate the propagation of embeddings. These GCN-inspired models represent the cutting edge in CF, underscoring the critical role of graph convolutional layers in modeling HOC.

However, this paper posits different perspectives that the integration of the contrastive loss function within any encoder can equivalently achieve HOC modeling, from the previous theoretical analysis, and that a careless use of the convolutional layer might lead to suboptimal results.

\subsection{High-Order Connectivity from Contrastive Loss}
Consider $\mathbf{E}(0)$ as the embeddings randomly initialized at step $0$.
By iteratively applying Equation (\ref{core}) for $T$ steps, we derive the embeddings at time step $T$
\begin{equation}
\label{stepT}
    \mathbf{E}(T) = \prod_{t=0}^{T-1} \left(\mathbf{I} + \gamma \mathbf{A}''(t)\right) \mathbf{E}(0),
\end{equation}
effectively stacking $T$ graph convolution operations.
To a certain extent, $\mathbf{E}(T)$ in the equation mirrors a $T$-layered GCN model—albeit devoid of linear transformations and nonlinear activations between layers. Notably, LightGCN \cite{he2020lightgcn} demonstrated that eliminating these elements enhances model performance in the CF context.
Typically, $T$ is considerably large, which implies the application of a multitude of graph convolution operations on the embeddings. Such operations foster message exchange between nodes \cite{gilmer2017neural}. With massive convolutions, not only is information from a node propagated to its high-order neighbors, but also the message exchange recurs until an equilibrium is attained.
If we posit that the embeddings effectively capture high-order connectivity, then it is reasonable to anticipate that the embedding of a single node can assimilate information from its high-order neighbors.
While traditional graph-based methods utilize convolutional layers to assimilate neighbor information, contrastive learning implicitly incorporates convolutions within its learning process.
Viewed from this angle, we advocate that contrastive learning bestows the capability of embeddings to model high-order connectivity.
\subsection{On The Necessity of Graph Convolutional layers}
Given that contrastive learning can effectively capture high-order connectivity, relevant questions arise: \textit{How does the combination of a contrastive objective and a graph-based model perform?} and \textit{Is there a genuine need for graph convolutional layer designs in CF?} To probe these concerns, we explore LightGCN \cite{he2020lightgcn}, a state-of-the-art graph-based model, with DirectAU \cite{wang2022towards}, a recently introduced contrastive objective directly optimizing embedding alignment and uniformity:
\begin{equation}
\label{eq_directau}
\begin{split}
l_{\text {align }} & =\mathbb{E}_{(u, i) \sim p_{\text {pos }}} \| \e_u -\e_i \|^2,\\
l_{\text {uniform }} & =\log \mathbb{E}_{u, u^{\prime} \sim p_{\text {user }}} e^{-2 \| \e_u - \e_{u'} \|^2} \\
& \quad + \log \mathbb{E}_{i, i^{\prime} \sim p_{\text {item }}} e^{-2 \| \e_i -\e_{i'} \|^2} ,\\
l_{\text{DirectAU}} &= l_{\text {align }} + \beta l_{\text {uniform }} 
\end{split}
\end{equation}

\subsubsection{A Brief Review of LightGCN}
The LightGCN model adopts the following graph convolution to propagate the embeddings:
\begin{equation}
\mathbf{E}^{(k+1)}=\left(\mathbf{D}^{-\frac{1}{2}} \mathbf{A} \mathbf{D}^{-\frac{1}{2}}\right) \mathbf{E}^{(k)},
\end{equation}
where $k$ is the number of layers and $\mathbf{E}^{(0)}$ is the learnable embeddings.
The final embedding for the prediction of a $K$-layered LightGCN is a weighted sum of each layer's embeddings
\begin{equation}
\label{lightgcn1}
    \mathbf{E} = \sum_{i=0}^K \alpha_i \Tilde{\mathbf{A}}\mathbf{E}^{(i)} = \left(\sum_{i=0}^K \alpha_i \Tilde{\mathbf{A}}^i \right)\mathbf{E}^{(0)},
\end{equation}
where $\Tilde{\mathbf{A}} = \mathbf{D}^{-1/2}\mathbf{A}\mathbf{D}^{-1/2}$. 

\subsubsection{Empirical Results}
\begin{table}[t!]
\small
\centering
\captionsetup{font=small}  
\caption{The performance of naive embedding and LightGCN with the DirectAU loss \cite{wang2022towards} on the Yelp2018 dataset. \# T.L. denotes the number of training layers and \# I.L. denotes the number of inference layers. No.S. denotes the experiment setting number.}
\label{directau_mf_lightgcn}
  {%
    \begin{tabular}{c|cc|cc}
    \toprule
    No. S. & \# T.L. & \# I.L. & Recall@20 & NDCG@20\\ \midrule
    0 & 0 & 0 & 0.1097 & 0.0684 \\
    1 & 1 & 1 & 0.1083 & 0.0677 \\
    2 & 2 & 2 & 0.1088 & 0.0686 \\
    3 & 3 & 3 & 0.1070 & 0.0675 \\
    4 & 3 & 0 & 0.0909 & 0.0564 \\
    5 & 3 & 1 & 0.1103 & 0.0691 \\
    6 & 3 & 2 & 0.1103 & 0.0692 \\
    \bottomrule
    \end{tabular}%
    }
\vspace{-2ex}
\end{table}

We compare the naive embedding model (LightGCN without linear filter) with the LightGCN model of different layers to examine their respective performance with the recently proposed contrastive loss DirectAU \cite{wang2022towards}, Equation (\ref{eq_directau}).
To investigate its impact, we manipulate the number of layers applied to the learnable embeddings during both the training and inference phases. Specifically, "training layers" refer to those utilized during the model training process, whereas 'inference layers' are applied to derive the results.

As depicted in Table \ref{directau_mf_lightgcn}, there are two groups of experiments: setting 0, 1, 2, and 3 are the first group -- they are in a consistent setting that the numbers of layers in both stages are the same.
Setting 3, 4, 5, and 6 are the second group -- they have the same number of layers in training but different number of layers in inference.
In the first group, setting 0, 1, 2, 3, the naive model (No.0) marginally outperforms consistently layered LightGCN setups (No.1, No.2, No.3). 
In the second group, comparing No.3 with No.5 and No.6, with exactly the same learnable embeddings, the 1-layered inference model outperforms the 3-layered one.
All these results indicate an increase in layers does not always enhance performance.
He et al. \cite{he2020lightgcn} reported similar observations on the Yelp2018 and Amazon-Book datasets with the BPR loss function.
We confirm this phenomenon in our experiments.

This gives rise to a pivotal question: Are graph convolutional layers indispensable for CF? 
Recent graph-based models for CF have largely been inspired by GNN \cite{wang2019neural, he2020lightgcn, mao2021ultragcn, wu2021self}.
One of the disadvantages of graph convolutional layers is that this operation is discrete, meaning its inability to regulate the augmentation of smoothness in the learning process.
This lack of control results in oversmoothing, smoother embedding but worse performance.
In contrast, contrastive learning provides a more controllable alternative by progressively updating while preventing collapse due to negative samples.
From this perspective, contrastive learning appears to be a more intuitive choice for modeling HOC.
Notably, we are not discouraging any design of encoder, but do warn about the misuse of graph convolutional layers.

\section{A simple and effective approach}
In preceding discussions, we presented a theoretical analysis underscoring the HOC modeling potential of contrastive loss and empirically elucidated the inherent risks associated with graph convolutional layers. To corroborate this theoretical understanding, we introduce Simple Contrastive Collaborative Filtering (SCCF), a model based on a naive embedding model, without embedding propagation, and a refined contrastive loss function.
One notable advantage of our method is its time complexity.
The time complexity of SCCF for obtaining a single node embedding is $ O(1)$, as it relies on a look-up table. In contrast, GNN-based models have a time complexity of \( O(d^l) \), where \( d \) represents the greatest node degree, and \( l \) denotes the number of layers. 
In the following section, we provide a detailed elaboration on the SCCF method.

Let $\mathcal{B}=\{(u_1, i_1), \dots, (u_m, i_m) \}$ be a collection of data sampled from $\mathcal{D}$ uniformly with batch size $m$ and $\mathcal{B}_N = \{(u_i, i_j)\;|\;i,j=1,\dots,m \}$ be the collection of all possible user-item pair of batch $\mathcal{B}$. 
We design the following learning objective:
\begin{align}
    \label{eq30}
    l &= -\frac{1}{m}\sum_{(u,i)\in \mathcal{B}}\log\left(\text{sim}(\e_u, \e_i) \right)\\ \nonumber 
    &+ \log\left(\frac{1}{m^2} \sum_{(u', i')\in \mathcal{B}_N} \text{sim}(\e_{u'}, \e_{i'})  \right).
\end{align}
The similarity function $\text{sim}(\cdot, \cdot)$ is defined as the following:
\begin{equation}
\label{eq_loss}
    \text{sim}(\e_u, \e_i) = \exp\left( \frac{\e_u^\top\e_i}{\tau||\e_u||_2 ||\e_i||_2} \right) + \exp\left(\frac{1}{\tau}\left( \frac{\e_u^\top\e_i}{||\e_u||_2 ||\e_i||_2} \right)^2\right),
\end{equation}
where $\tau$ is identified as the temperature parameter \cite{chen2020simple}, and $||\cdot||_2$ represents the $L^2$ norm. Intriguingly, Equation (\ref{eq_loss}) can be interpreted as a mixture of two exponential kernels \cite{tan2023contrastive}. It is pivotal to pinpoint some nuances differentiating Equation (\ref{infonce}) from Equation (\ref{eq_loss}): first, the introduction of the temperature parameter $\tau$; second, the shift from the inner product to cosine similarity; third, the incorporation of a second-order cosine similarity to transcend mere linearity.

First, the integration of the temperature parameter has been empirically shown to be crucial \cite{chen2023adap, chen2020simple, he2020momentum} as it modulates the relative disparities between samples \cite{wang2021understanding}. 
Equation (\ref{infonce}) can be interpreted as a special case where $\tau$ equals $1$.
Second, the cosine similarity can be considered as the inner product between embeddings with $L^2$ normalization.
This normalization not only acts as a regularization but also imposes constraints on gradients \cite{chen2023adap}. Moreover, the $L^2$ normalization establishes a link between the inner product and the Mean Squared Error (MSE), which is evident from the relationship: $(\bm{x}-\bm{y})^2 = 2 - \bm{x}^\top\bm{y}$ for $||\bm{x}||_2=||\bm{y}||_2=1$. 
Given the softmax function's invariant nature \cite{Goodfellow-et-al-2016}, the normalized inner product is congruent with MSE. This alignment relates our contrastive objective with the radial basis function kernel \cite{hofmann2008kernel}.
Third, the inclusion of the second-order cosine similarity aims to infuse non-linearity into our learning objective. As substantiated by \cite{tan2023contrastive}, the mixture of varied kernels (similarity functions) augments performance.

\begin{table}
\small
\centering
\captionsetup{font=small}
\caption{Statistics of datasets.}
\begin{tabular}{@{}lrrrr@{}}
\toprule
Dataset       & \# Users & \# Items & \# Interactions & Sparsity \\ \midrule
Amazon-Beauty & 22,363   & 12,101   & 198,502         & 99.93\%  \\
Gowalla       & 29,858   & 40,891   & 1,027,370        & 99.92\%  \\
Yelp2018      & 31,668   & 38,048   & 1,561,406        & 99.87\%  \\
Pinterest     & 55,187   & 9,912    & 1,445,622        & 99.74\%  \\ \bottomrule
\end{tabular}
\label{table: datasets}
\vspace{-2ex}
\end{table}

\section{Experiments}\begin{table*}[h]
\small
    \centering
    \captionsetup{font=small} 
    \caption{Comparison between baselines and our method. The $\ddagger$ denotes significant improvements with t-test at $p<0.05$ over all compared methods. The best results are in \text{bold} and the second best results are \underline{underlined}.}
    \label{tab_3}
    \begin{tabular}{ccccccccccc|c}
    
    \toprule
         Dataset&  Metric&  BPR&  DAU&  LGCN-B&  LGCN-D&  Mult-VAE&  SimpleX&  NGCF & DGCF & SGL &  SCCF\\
         \midrule
         Beauty&  Recall@20&  0.1210&  0.1344&  0.1249&  \underline{0.1437}&  0.1114&  0.1188& 0.1072 & 0.1142 & 0.1387 &  \textbf{0.1470}$^\ddagger$ \\
         &  NDCG@20&  0.0564&  0.0665&  0.0591&  \underline{0.0683}&  0.0541&  0.0550& 0.0490 & 0.0538 & 0.0681 & \textbf{0.0713}$^\ddagger$\\
         \midrule
         Gowalla&  Recall@20&  0.1303&  0.2020&  0.1914&  0.1994&  0.1775&  0.1114& 0.1580 & 0.1825 & \underline{0.2139} & \textbf{0.2185}$^\ddagger$ \\
         &  NDCG@20&  0.0771&  0.1145&  0.1103&  0.1136&  0.1008&  0.0557& 0.0904 & 0.1076 & \underline{0.1271} & \textbf{0.1304}$^\ddagger$ \\
         \midrule
         Yelp2018&  Recall@20&  0.0612&  \underline{0.1097}&  0.0896&  \text{0.1070}&  0.0922&  0.0715& 0.0808 & 0.0852 & 0.1064 & \textbf{0.1160}$^\ddagger$ \\
         &  NDCG@20&  0.0375&  0.0684&  0.0550&  \underline{0.0675} &  0.0559&  0.0422& 0.0485 & 0.0527 & 0.0669 & \textbf{0.0728}$^\ddagger$ \\
         \midrule
         Pinterest&  Recall@20&  0.1278 &  0.1477&  0.1625&  0.1764&  0.1692&  0.1376& 0.1334 & 0.1571 & \underline{0.1775} & \textbf{0.1776} \\
         &  NDCG@20&  0.0645&  0.0834&  0.0840&  0.0921&  \underline{0.1010}&  0.0666& 0.0675 & 0.0807 & 0.0952 & \textbf{0.1035}$^\ddagger$\\
         \bottomrule
    \end{tabular}
\vspace{-2ex}
\end{table*}

This section is organized as follows:
Initially, the experimental settings are given. 
Subsequently, we show that our proposed model, SCCF, demonstrates equivalent or superior performance in comparison to several state-of-the-art methods. 
Later, we present evidence indicating that the incorporation of graph convolution may lead to a suboptimal performance.
Lastly, we conduct ablation studies to analyze the impact of various components in SCCF.
\subsection{Experimental Settings}
\subsubsection{Datasets}
We utilize four real-world datasets for our experiments: \textbf{Amazon-Beauty}\footnote{https://cseweb.ucsd.edu/~jmcauley/datasets.html\#amazon\_reviews} comprises users' online shopping records on the Amazon website; 
\textbf{Gowalla}\footnote{http://snap.stanford.edu/data/loc-gowalla.html} consists of users' check-in information from a social networking website;
\textbf{Yelp2018}\footnote{https://www.yelp.com/dataset} includes information about businesses, reviews, and user data for academic purposes; 
\textbf{Pinterest}\footnote{https://sites.google.com/site/xueatalphabeta/academic-projects} originally is proposed in \cite{geng2015learning} and later adopted by \cite{he2017neural} for image recommendation.
Dataset statistic information is provided in Table~\ref{table: datasets}.

\subsubsection{Evaluation Protocols}
For each dataset, we randomly split each user's interactions into training/validation/test sets with a ratio of 80\%/10\%/10\%.
For the evaluation of Top-K recommendation performance, we employ two metrics: Recall and Normalized Discounted Cumulative Gain (NDCG).
Recall@K assesses whether the test ground-truth items are present in the retrieved Top-K list.
NDCG@K evaluates the position of the ground-truth items in the Top-K list, considering the relevance and rank positions.

\subsubsection{Implementation Detail}
To ensure fair comparisons, we employed the RecBole framework \cite{zhao2021recbole} across all experimental methods. The embedding size was fixed at 64 and initialized using Xavier normal initialization for all models. Training epochs were set to 300, from which the model with the best performance on the validation set was selected. For the SCCF method, we used naive SGD as the optimizer. The batch sizes were set to 10,000, 100,000, 100,000, and 60,000 for the Beauty, Gowalla, Yelp2018, and Pinterest datasets, respectively. The temperature parameter $\tau$ was set to 0.25, 0.1, 0.2, and 0.1 for the Beauty, Gowalla, Yelp2018, and Pinterest datasets, respectively.

\subsubsection{Baselines} We compare our method with several baseline algorithms, including: (1) \textbf{BPR} \cite{rendle2012bpr}: The naive embedding model with BPR loss, where the negative item is randomly sampled from the item set. (2) \textbf{DAU} \cite{wang2022towards}: The naive embedding model incorporates the DirectAU loss, which directly optimizes alignment and uniformity. (3) \textbf{LGCN-B} \cite{he2020lightgcn}: The LightGCN model is with the BPR loss from the original paper. (4) \textbf{LGCN-D} \cite{wang2022towards}: The LightGCN model combines with the DirectAU loss. (5) \textbf{Mult-VAE} \cite{liang2018variational}: A variational autodecoder beyond linear models. (6) \textbf{SimpleX} \cite{mao2021simplex}: A simple but strong baseline consisted of a SVD++-like model and cosine contrastive loss. (7) \textbf{NGCF} \cite{wang2019neural}: This model employs a GCN-like architecture with learnable linear transformations and non-linear activation functions.
(8) \textbf{DGCF} \cite{wang2020disentangled}: Focusing on modeling diverse relationships, This graph-based model focuses on modeling diverse user intent disentanglement.
(9) \textbf{SGL} \cite{wu2021self}: This method introduces three data augmentations for graph-based CF models.

\subsection{Comparison with Other Baselines}
Table \ref{tab_3} provides a comparison of the different models and loss functions in the metric of Recall@20 and NDCG@20. A comprehensive table, Table \ref{comparison}, is provided in the Appendix for a detailed comparison.
Across all evaluated datasets, our method, SCCF, consistently outperforms competing approaches, achieving the best performance. Among the alternative models, LGCN-D and SGL demonstrate substantial efficacy. Notably, despite SGL's application of three data augmentation techniques—edge drop, node drop, and random walk—to bolster training, our SCCF model, without any augmentations, still manages to exceed SGL's performance. This outcome underscores the effectiveness of the SCCF approach.

Furthermore, we can categorize the evaluated methods into two distinct groups. The first group, comprising BPR, DAU, and SCCF, implements a naive embedding model with specifically designed loss functions. The second group includes all graph-based methods: LGCN-B, LGCN-D, NGCF, DGCF, SGL, and SCCF.
Within the first group, the primary distinction lies in the design of the loss function. Our proposed loss functions, Equation (\ref{eq30}), and DirectAU's loss, Equation (\ref{eq_directau}), differentiate themselves by incorporating multiple negative samples, in contrast to BPR's single negative sample approach. This multiplicity of negative samples may explain the superior performance of our and DirectAU's loss functions over BPR's. Specifically, our loss function utilizes user-item pairs as negative samples, whereas DirectAU employs user-user and item-item pairs. This distinction suggests that our method, by directly expanding the distances between user and item embeddings, may offer advantages over DirectAU, which indirectly achieves this objective by expanding distances between user-user and item-item. These observations highlight the design of the loss function.

In the second group, all graph-based models implement graph convolutional layers as a fundamental component. In contrast, our SCCF model eschews graph convolutional layers yet achieves superior performance across these graph-based approaches. This observation suggests that contrastive loss provides a more adaptable mechanism for embedding propagation, potentially surpassing the capabilities of graph convolutional layers.
Furthermore, considering the prevalent assumption that graph convolutional layers are effective in modeling HOC, the outperformance of SCCF invites a reevaluation of this premise. Specifically, if SCCF, which relies on contrastive loss, surpasses graph-based models in performance, it implies that contrastive loss is either equally capable of modeling HOC or challenges the notion that HOC is advantageous for CF tasks.
The findings from this group indicate that graph convolutional layers are not indispensable for HOC modeling. Instead, they highlight the efficacy of contrastive loss in achieving, and potentially exceeding, the modeling capabilities attributed to graph convolutions.

\subsection{Comparison between Naive Embedding and LightGCN}
Although the preceding section demonstrates the effectiveness of SCCF by comparing it with various graph-based models, this comparison might not be entirely fair, given that different models may employ distinct loss functions.
To more accurately assess the influence of graph convolutional layers on performance, we conduct a comparative analysis employing various encoders (LightGCN with different layers and the naive embedding model) under a consistent contrastive loss function, Equation (\ref{eq30}).
As Table \ref{mf_ligtgcn} shows, the naive model outperforms all LightGCN models, with significant improvement on the Gowalla and Yelp2018 datasets.
A more comprehensive Table \ref{tab_mf_light} is provided in the Appendix.
One reason to account for this phenomenon is that the graph convolutional layer introduces an inductive bias conducive to modeling HOC. A weak loss function, such as the BPR loss, may result in embeddings that lack sufficient concentration; in such contexts, the inductive bias introduced by graph convolution can be beneficial in enhancing focus. Conversely, when the model is equipped with a contrastive loss, embeddings tend to be well-concentrated, rendering the additional inductive bias unnecessary and increasing the risk of oversmoothing.

\begin{table}[h]
\small
\centering
\captionsetup{font=small}  
\caption{Effectiveness of different encoders. The \textbf{bold} denotes the best result. "LGCN-x" denotes LightGCN with x layers. "NE" denotes the naive embedding model without any convolutional layer.}
\scalebox{1.0}{
\begin{tabular}{*{6}{c}}
  \toprule
  Dataset & Metric & LGCN-1 & LGCN-2 & LGCN-3 & NE \\
  \midrule
 \multirow{2}*{Beauty} & Recall@20 & 0.1418 & 0.1437 & 0.1437 & \textbf{0.1470}\\
  & NDCG@20 & 0.0669 & 0.0685 & 0.0684 & \textbf{0.0713}\\
  \midrule
  \multirow{2}*{Gowalla}& Recall@20 & 0.2020 & 0.2065 & 0.2004 & \textbf{0.2185} \\
  & NDCG@20 & 0.1168 & 0.1203 & 0.1169 & \textbf{0.1304}\\
  \midrule
 \multirow{2}*{Yelp2018} & Recall@20 & 0.1062 & 0.1079 & 0.1062 & \textbf{0.1160}\\
  & NDCG@20 & 0.0665 & 0.0679 & 0.0667 & \textbf{0.0728}\\
  \midrule
  \multirow{2}*{Pinterest}& Recall@20 & 0.1712 & 0.1768 & 0.1676 & \textbf{0.1776}\\
  & NDCG@20 & 0.0960 & 0.0992 & 0.0918 & \textbf{0.1035}\\
  \bottomrule
\end{tabular}}
\label{mf_ligtgcn}
\vspace{-2ex}
\end{table}

\subsection{Hyperparameter and Ablation Study}
\subsubsection{Impact of $L^2$ normalization}
\begin{table}[t]
\small
\centering 
\captionsetup{font=small}
\caption{Comparison of various similarity functions during the training and inference stages on two datasets. "Cos" denotes the cosine similarity and "IP" denotes the inner product.}
\label{train_inference}
\begin{tabular}{ccc|cc}
\toprule
Dataset & Training & Inference & Recall@20 & NDCG@20 \\ \midrule
\multirow{4}{*}{Beauty} &
Cos & Cos & 0.1127 & 0.0551\\
&Cos & IP & 0.1470& 0.0713\\
&IP & Cos & 0.0390 & 0.0171 \\
&IP & IP & 0.0773 & 0.0370 \\ \midrule
\multirow{4}{*}{Gowalla} &
Cos & Cos & 0.1501 & 0.0785\\
&Cos & IP & 0.2185 & 0.1304\\
&IP & Cos & 0.0335 & 0.0163 \\
&IP & IP & 0.1487 & 0.0805 \\ \bottomrule
\end{tabular}
\end{table}
Table~\ref{train_inference} displays the effectiveness of various similarities during the training and inference stages. Remarkably, employing cosine similarity during training and the inner product during inference yields the most superior results. The optimization using cosine similarity in training is more challenging than the inner product because it disregards the magnitude of the embeddings. Practically speaking, popular items or active users generally exhibit larger magnitudes and subsequently have higher scores than their inactive counterparts. By disregarding magnitude, we effectively mitigate popularity bias, enabling the mining of patterns beyond mere frequency. Conversely, during the inference stage, the magnitude of embeddings becomes pivotal as it reflects the popularity of user/item, playing a crucial role in recommendations.

\subsubsection{Impact of temperature}
\begin{figure}
\small
\centering
\includegraphics[width=0.7\linewidth]{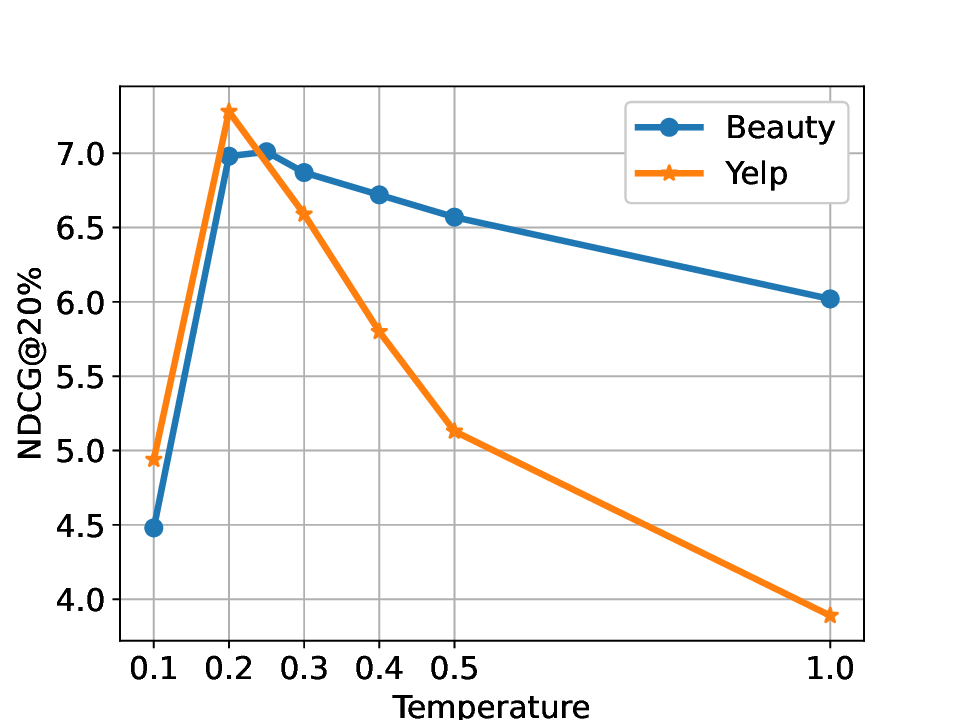}
\captionsetup{font=small}
\caption{Performance with different temperatures on Amazon-Beauty and Yelp2018 dataset.}
\label{temperature}
\end{figure}
As illustrated in Figure \ref{temperature}, the performance of the model exhibits variation with respect to the temperature parameter. The temperature parameter $\tau$ controls the smoothness of the similarity distribution, thereby regulating the impact of negative samples. A smaller value of $\tau$ makes the model more sensitive to hard negative samples, as they contribute significantly to the loss. Conversely, as $\tau$ increases, the model becomes less sensitive to individual samples and focuses more on the overall distribution.
Noticing that the optimal value of $\tau$ may vary across datasets, indicates the importance of selecting the appropriate temperature based on the specific characteristics of the dataset being used.

\subsubsection{Impact of embedding size}
\begin{figure}
\small
    \centering
    \begin{minipage}{0.45\linewidth}
        \includegraphics[width=\linewidth]{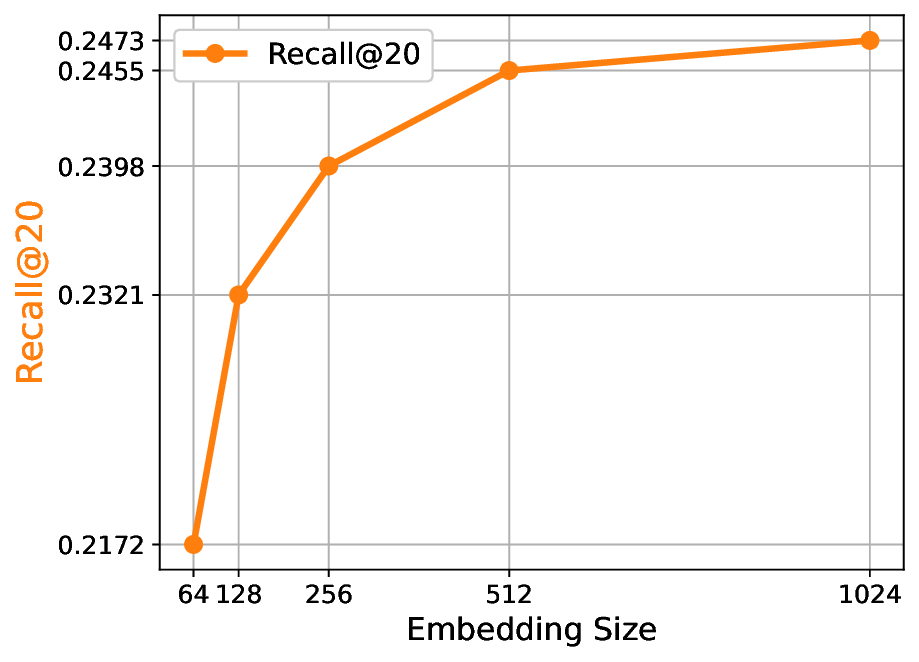}
    \end{minipage}
    \hfill
    \begin{minipage}{0.45\linewidth}
        \includegraphics[width=\linewidth]{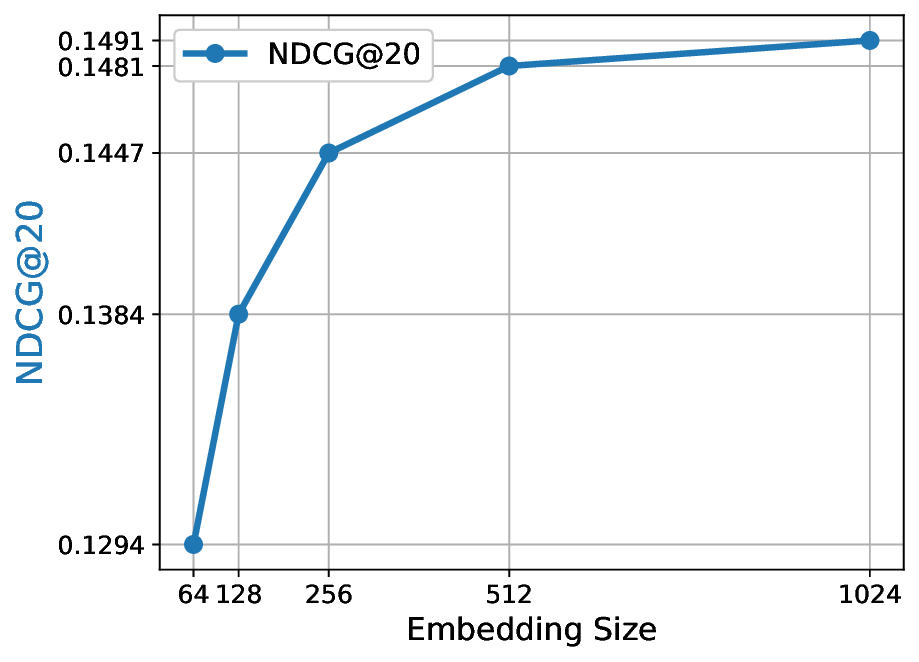}
    \end{minipage}
    \captionsetup{font=small}
    \caption{Embedding size vs. Recall@20 (left) and NDCG@20 (right) on the Gowalla dataset.}
    \label{fig_embsize}
\end{figure}
As Figure \ref{fig_embsize} shows, the model's performance improves with the increase of embedding dimension and is saturated when the dimension reaches a certain degree, which aligns with our common sense that more parameters, better performance.
The preference for lower dimensions can be attributed not only to computational efficiency but also to early works on MF like Funk's SVD\footnote{https://sifter.org/~simon/journal/20061211.html} and SVD++ \cite{koren2008factorization}, which essentially are low-rank approximations of the interaction matrix.
The foundational assumption of these methods is that low-rank decomposition, specifically those leading singular values and their corresponding singular vectors, suffices to approximate the original matrix accurately.
At first glance, our results might appear to challenge this low-rank paradigm.
One hypothesis proposed to explain these phenomena is the independence assumption \cite{levy2014neural}, which posits that only sufficiently large dimensionalities can effectively decouple the correlations among different user-item pairs.
The violation of the independence assumption might lead to an inaccuracy modeling of the empirical distribution.
In other words, a low-dimension embedding may not be able to characterize the complex constraints on embeddings, i.e., the complex interaction between users and items, explaining the suboptimal performance with low-dimension.

\subsubsection{Impact of square of cosine similarity}
\begin{table}
\small
\centering
\captionsetup{font=small}
\caption{The effectiveness of the square of cosine similarity on two datasets, where "w/" and "w/o" denote the model with or without the second-order similarity. }
\label{tab_second_order}
    \begin{tabular}{c|cc|cc}
    \toprule
         &  \multicolumn{2}{c}{Beauty} & \multicolumn{2}{|c}{Pinterest}\\ 
         &  Recall@20&  NDCG@20&  Recall@20& NDCG@20\\
         \midrule
         w/o&  0.1419&  0.0701&  0.1734& 0.1015\\ 
         w/&  0.1470&  0.0713&  0.1776& 0.1035\\
         \bottomrule
    \end{tabular}
\vspace{-2ex}
\end{table}
In Table \ref{tab_second_order}, it is evident that integrating the square of cosine similarity enhances the model's performance. While the primary performance can be attributed to the first-order cosine similarity, the squared term still provides a notable improvement. We postulate that this enhancement arises from the second-order interaction between embeddings, thereby facilitating a more refined representation.

\section{Related Work}
\textbf{Collaborative Filtering}.
Collaborative Filtering (CF) is a fundamental and important algorithm for recommender systems.
Model-based methods have gained more popularity over memory-based methods since the latter often relies on heuristics.
A notable example of model-based methods is Matrix Factorization \cite{koren2009matrix}, which decomposes the interaction matrix into two low-dimensional matrices.
NeuMF \cite{he2017neural} and Multi-VAE \cite{liang2018variational} are two representative non-linear methods for CF.
Recently, GNN-based embedding models have gained much attention in CF.
Inspired by GCN \cite{kipf2016semi}, NGCF \cite{wang2019neural} utilizes graph convolutional layers to propagate embeddings. LightGCN \cite{he2020lightgcn} removes linear transformations and non-linear activations in graph convolutional layers to improve performances. Additionally, \citet{shen2021powerful} provided a unified analysis through low-pass filtering, and proposed an effective graph filtering model, GF-CF.
Existing GNN-based models in CF employ graph convolutional layers to capture high-order connectivity. Nonetheless, our findings reveal naive embedding with a contrastive objective demonstrates comparable capability in modeling such connectivity. \\

\noindent \textbf{Contrastive Learning in CF}.
The BPR \cite{rendle2012bpr} loss function is a pioneer of contrastive learning in CF.
CLRec \cite{zhou2021contrastive} uses the contrastive loss function to reduce exposure bias through inverse propensity weighting.
Drawing insights from BYOL \cite{grill2020bootstrap}, BUIR \cite{lee2021bootstrapping} incorporates a momentum update, enabling the training of embedding encoders devoid of negative samples.
\citet{wang2020understanding} identified two properties for contrastive learning, the alignment for positive samples and the uniformity of features' distribution.
Inspired by this idea, \citet{wang2022towards} proposed the DirectAU loss as a means to enhance the alignment
and uniformity of embeddings in CF. \\

\noindent \textbf{Graph Contrastive Learning in CF}.
Another significant topic is graph contrastive learning, which aims to learn invariant representations through graph perturbation. SGL \cite{wu2021self} proposed three graph augmentation methods: node dropout, edge dropout, and random walk. SimGCL \cite{yu2022graph} introduced the idea of adding uniform noise to node embeddings. LightGCL \cite{cai2023lightgcl} aligned node embeddings with their SVD-augmented counterparts.
However, all of these methods rely on graph convolutional layers to implement graph augmentation. For simplicity and generality, our discussion on the contrastive objective does not involve any graph augmentation and remains independent of specific models. The investigation of model-agnostic augmentation methods is left for future work. \\

\noindent \textbf{Theory of Contrastive Learning}.
\citet{haochen2021provable} elucidated the role of contrastive learning in performing spectral clustering on the augmentation graph. Taking the concept of augmentation overlap into account, \citet{wang2022chaos} illustrated how aligned data augmentations facilitate the clustering of intra-class samples. \citet{wang2023message} proposed that the learning dynamics of contrastive learning resonates with message-passing mechanisms on the augmentation graphs and affinity graph.
While our analysis bears similarities to that of \citet{wang2023message}, there are distinct differences in focus. \citet{wang2023message} concentrates on general contrastive learning, incorporating data augmentation techniques. In contrast, our research specifically targets the CF setting, eschewing data augmentation in favor of a deeper exploration of graph theory.

\section{Conclusion}
In this study, we reexamine graph convolution and contrastive learning in the context of collaborative filtering and reveal the equivalence between them.
This equivalence offers a new perspective to analyze contrastive learning via graph theory. By doing so, we show the capacity of contrastive learning for high-order connectivity modeling.
Moreover, we examine whether it is necessary to add graph convolutional layers to model high-order connectivity. We show that this is unnecessary. 

Based on the above analysis, we propose a simple and effective algorithm using a new contrastive loss, enabling the model to produce equivalent or even superior performance compared with other graph-based methods. This further confirms that a model using contrastive loss can successfully capture high-order connectivity, which was believed to be obtained only with graph convolution.

This paper is a first step in trying to better understand CF algorithms using graph convolution and contrastive learning. We believe the problem should be further investigated to gain more insight into the models for designing better CF algorithms.

\begin{acks}
    This work was partly supported by the NSERC discovery grant and the Canada Research Chair on natural language information processing and applications.
\end{acks}



\bibliographystyle{ACM-Reference-Format}
\balance
\bibliography{ref}

\appendix

\section{Proof for Theorem \ref{theorem1}}
\label{proof_th1}
\newcommand{\D}{\mathbf{D}}
\newcommand{\x}{\bm{x}}
The core of proof  lies in the inequality
\begin{equation}
\label{eq26}
    S\left((\mathbf{I}+\mathbf{D})\bm{x}\right) < \frac{\|\mathbf{I}+\D_{\text{max}} \|^2}{\| \mathbf{I}+ \D\|^2}S(\bm{x}),
\end{equation}
where $\bm{x}$ is the graph signal, $\D$ is the degree matrix, $\D_{\text{max}}$ is a diagonal matrix filled by the greatest degree.
The key observation is that $S((\mathbf{I}+\gamma \mathbf{A}/|\mathcal{D}|)\bm{x})$ is bounded by $S((\mathbf{I}-\gamma \mathbf{L}/|\mathcal{D}|)\bm{x})$, a low-pass filter; with small enough learning rate $\gamma$, eventually $S((\mathbf{I}-\gamma \mathbf{L}/|\D|)\bm{x}) \ll S(\bm{x})$.
Let's prove Equation (\ref{eq26}) first:
\begin{align*}
    &S((\mathbf{I}+\mathbf{D})\bm{x}) = \frac{\x^\top\x+\x^\top\D\mathbf{L}\x+\x^\top\mathbf{L}\D\x+\x^\top\D\mathbf{L}\D\x}{\|\mathbf{I}+\D \|^2 \x^\top\x}\\
    &\leq \frac{\x^\top\x+\x^\top\D_{\text{max}}\mathbf{L}\x+\x^\top\mathbf{L}\D_{\text{max}}\x+\x^\top\D_{\text{max}}\mathbf{L}\D_{\text{max}}\x}{\|\mathbf{I}+\D \|^2 \x^\top\x}\\
    &= \frac{\|\mathbf{I}+\D_{\text{max}} \|^2}{\| \mathbf{I}+ \D\|^2}S((\mathbf{I}+\D_{\text{max}})\bm{x})\\
    &= \frac{\|\mathbf{I}+\D_{\text{max}} \|^2}{\| \mathbf{I}+ \D\|^2}S(\bm{x})
\end{align*}
The first and the third equality are obtained by the definition of $S(\cdot)$.
The less than sign is obtained by the fact that $(\D_{\text{max}}-\D)\mathbf{L}$ is a semi-positive matrix.
The last equality is due to the fact that $S(\bm{x})=S((\mathbf{I}+\D_{\text{max}})\bm{x})$.
Replacing $\mathbf{I}$ with $\mathbf{I}-\gamma \mathbf{L}/|\mathcal{D}|$ and $\D$ with $\gamma \D/|\mathcal{D}|$ in Equation (\ref{eq26}), we have
\begin{equation}
\label{eq27}
    S((\mathbf{I}+\gamma \mathbf{A}/|\D|)\bm{x}) \leq \frac{\|\mathbf{I}-\gamma \mathbf{L}/|\mathcal{D}| + \gamma \D/|\mathcal{D}| \|^2}{\|\mathbf{I}+\gamma \mathbf{A}/|\mathcal{D}| \|^2}S((\mathbf{I}-\gamma \mathbf{L}/|\mathcal{D}|)\x).
\end{equation}
By Proposition \ref{prop2} and Equation (\ref{eq27}), we can obtain small enough $\gamma$ so that $S((\mathbf{I}+\gamma \mathbf{A}/|\D|)\bm{x}) < S(\x)$: $\mathbf{I}+\gamma \mathbf{A}/|\D|$ increase signal's smoothness.
The proof of $\mathbf{I}-\gamma \mathbf{A}'(t)$ can be obtained in a similar way.

\section{Proof for Theorem \ref{theorem2}}
\label{proof_th2}
When the system stabilizes into equilibrium, it adheres to the condition:
    $\mathbf{E}(\infty) = \mathbf{E}(\infty) + \gamma\mathbf{A}''\mathbf{E}(\infty)$.
It follows that $\mathbf{A}''\mathbf{E}(\infty) = \mathbf{0}$.
Since $\mathbf{E}$ cannot be a zero matrix $\mathbf{0}$, either $\mathbf{A}''$ is $\mathbf{0}$ or $\mathbf{E}$ is a non-zero solution for $\mathbf{A}''\mathbf{E}=\mathbf{0}$ when $\mathbf{A}'' \neq \mathbf{0}$.
We will demonstrate that in the second case, the non-zero solution \(\mathbf{E}\) is unstable; any perturbation at this stationary point causes the system to move away from it.
Consequently, the embedding system by the contrastive objective reaches its equilibrium if and only if $\mathbf{A}''=\mathbf{0}$.

To determine the stability of the stationary points, we turn to the knowledge of dynamical systems \cite{verhulst2006nonlinear}.
When $d \mathbf{E}/dt = \mathbf{A}''\mathbf{E} = \mathbf{0}$, the stability of $\mathbf{E}$ is determined by $\mathbf{A}''$: the embeddings are stable if and only if all the eigenvalues of $\mathbf{A}''$ are equal to or less than $0$.
Since our graph has no self-loop, the diagonal of $\mathbf{A}''$ are zeros.
This zero trace indicates two cases: (1) all the eigenvalues are zeros; (2) there exist at least one positive and one negative eigenvalues.
Since $\mathbf{A}''$ is symmetric, $\mathbf{A}''$ is not defective; i.e., it has full eigenvectors corresponding to its number of rows.
It is impossible for $\mathbf{A}''$ to have all zero eigenvalues unless $\mathbf{A}'' = \mathbf{0}$, which contradicts our assumption that $\mathbf{A}'' \neq \mathbf{0}$.
Therefore, there must be some eigenvalues that are non-zero.
Recall the fact that those non-zero eigenvalues must sum up to zero, then at least there exists one positive eigenvalue which makes the stationary point unstable.

\section{Tables}
Tables \ref{tab_mf_light} and \ref{comparison} provide detailed experimental results with additional metrics.
\begin{table}[h!]
\centering
\caption{Performance of different models on four datasets. The \textbf{bold} denotes the best result. "NE" denotes naive embeddings.}
\scalebox{0.95}{
\begin{tabular}{*{6}{c}}
  \toprule
  Dataset & Metric  & LGCN-1 & LGCN-2 & LGCN-3 & NE  \\
  \midrule
  \multirow{6}*{Beauty} & Recall@10  & 0.0982 & 0.1028 & 0.1026 & \textbf{0.1060} \\
  & Recall@20  & 0.1418 & 0.1437 & 0.1437 & \textbf{0.1470}\\
  & Recall@50  & 0.2089 & 0.2130 & \textbf{0.2143} & 0.2132 \\
  \cmidrule(lr){2-6}
  & NDCG@10  & 0.0555 & 0.0579 & 0.0577 & \textbf{0.606}\\
  & NDCG@20  & 0.0669 & 0.0685 & 0.0684 & \textbf{0.0713}\\
  & NDCG@50  & 0.0806 & 0.0827 & 0.0828 & \textbf{0.0849} \\
  \midrule
  \multirow{6}*{Gowalla} & Recall@10  & 0.1394 & 0.1422 & 0.1382 & \textbf{0.1552} \\
  & Recall@20  & 0.2020 & 0.2065 & 0.2004 & \textbf{0.2185} \\
  & Recall@50  & 0.3162 & 0.3203 & 0.3157 & \textbf{0.3318} \\
  \cmidrule(lr){2-6}
  & NDCG@10  & 0.0987 & 0.1017 & 0.0989 & \textbf{0.1122} \\
  & NDCG@20  & 0.1168 & 0.1203 & 0.1169 & \textbf{0.1304}\\
  & NDCG@50  & 0.1448 & 0.1482 & 0.1450 & \textbf{0.1581}\\
  \midrule
  \multirow{6}*{Yelp2018} & Recall@10  & 0.0656 & 0.0668 & 0.0659 & \textbf{0.0726} \\
  & Recall@20  & 0.1062 & 0.1079 & 0.1062 & \textbf{0.1160}\\
  & Recall@50  & 0.1923 & 0.1944 & 0.1920 & \textbf{0.2042}\\
  \cmidrule(lr){2-6}
  & NDCG@10  & 0.0530 & 0.0542 & 0.0532 & \textbf{0.0583}\\
  & NDCG@20  & 0.0665 & 0.0679 & 0.0667 & 0.\textbf{0728}\\
  & NDCG@50  & 0.0915 & 0.0928 & 0.0914 & 0.\textbf{0982}\\
  \midrule
  \multirow{6}*{Pinterest} & Recall@10  & 0.1136 & 0.1175 & 0.1092 & \textbf{0.1200} \\
  & Recall@20  & 0.1712 & 0.1768 & 0.1676 & \textbf{0.1776}\\
  & Recall@50  & 0.2805 & \textbf{0.2922} & 0.2766 & 0.2911\\
  \cmidrule(lr){2-6}
  & NDCG@10  & 0.0778 & 0.0778 & 0.0734 & \textbf{0.0853}\\
  & NDCG@20  & 0.0960 & 0.0992 & 0.0918 & \textbf{0.1035}\\
  & NDCG@50  & 0.1232 & 0.1278 & 0.1189 & \textbf{0.1313}\\
  \bottomrule
\end{tabular}}
\label{tab_mf_light}
\end{table}

\begin{table*}
\caption{Comparison of different systems performance on four datasets. The $\ddagger$ denotes significant improvements with t-test at $p<0.05$ over all compared methods. The \textbf{bold} denotes the best results. The second best results are underlined.}
\label{comparison}
\centering
\begin{tabular}{ccccccccccc|c}
  \toprule
  Dataset & Metric & BPR & DAU & LGCN-B & LGCN-D & Mult-VAE & SimpleX & NGCF & DGCF & SGL & SCCF \\
  \midrule
  \multirow{6}*{Beauty} & Recall@10 & 0.0832 & 0.0973 & 0.0876 & \underline{0.1004} & 0.0784 & 0.0831 & 0.0743	& 0.0776 &	0.0992 & \textbf{0.1060}$^\ddagger$  \\
  & Recall@20 & 0.1210 & 0.1344 & 0.1249 & \underline{0.1437} & 0.1114 & 0.1188 & 0.1072	&0.1142	&0.1387&\textbf{0.1470}$^\ddagger$ \\
  & Recall@50 & 0.1831 & 0.1922 & 0.1887 & \underline{0.2113} & 0.1676 & 0.1833 & 0.1671&	0.1722	&0.2008&\textbf{0.2132} \\
  \cmidrule(lr){2-12}
  & NDCG@10 & 0.0466 & 0.0568 & 0.0494 & 0.0571 & 0.0445 & 0.0457 & 0.0404&	0.0443	&\underline{0.0578} &\textbf{0.0606}$^\ddagger$ \\
  & NDCG@20 & 0.0564 & 0.0665 & 0.0591 & \underline{0.0683} & 0.0541 & 0.0550 & 0.0490&	0.0538	&0.0681&\textbf{0.0713}$^\ddagger$ \\
  & NDCG@50 & 0.0691 & 0.0784 & 0.0721 & \underline{0.0820} & 0.0656 & 0.0683 & 0.0612&	0.0658	&0.0809&\textbf{0.0849}$^\ddagger$  \\
  \midrule
  \multirow{6}*{Gowalla} & Recall@10 & 0.0896 & 0.1360 & 0.1328 & 0.1370 & 0.1226 & 0.0637 & 0.1083&	0.1265	&\underline{0.1494}&\textbf{0.1552}$^\ddagger$ \\
  & Recall@20 & 0.1303& 0.2020 & 0.1914 & 0.1994& 0.1775& 0.1114 & 0.1580&	0.1825	&\underline{0.2139}&\textbf{0.2185}$^\ddagger$ \\
  & Recall@50 &     0.2078& 0.2989 & 0.2986 & 0.3103 & 0.2827 & 0.2056 & 0.2561&	0.2877&	\underline{0.3253}&\textbf{0.3318}$^\ddagger$ \\
  \cmidrule(lr){2-12}
  & NDCG@10 & 0.0655 & 0.0975 & 0.0934 & 0.0955 & 0.0849 & 0.0427& 0.0759&	0.0915	&\underline{0.1087}&\textbf{0.1122}$^\ddagger$ \\
  & NDCG@20 & 0.0771 & 0.1145 & 0.1103 & 0.1136 & 0.1008 & 0.0557 &0.0904	&0.1076	&\underline{0.1271} &\textbf{0.1304}$^\ddagger$ \\
  & NDCG@50 & 0.0961 & 0.1400 & 0.1365 & 0.1407 & 0.1263 & 0.0785 & 0.1143&	0.1332	&\underline{0.1544}&\textbf{0.1581}$^\ddagger$ \\
  \midrule
  \multirow{6}*{Yelp2018} & Recall@10 & 0.0364 & \underline{0.0685} & 0.0549 & 0.0670 & 0.0563 & 0.0412 & 0.0480&	0.0515	&0.0666&\textbf{0.0726}$^\ddagger$ \\
  & Recall@20 & 0.0612 & \underline{0.1097} & 0.0896 & 0.1070 & 0.0922 & 0.0715 &0.0808	&0.0852&	0.1064 &\textbf{0.1160}$^\ddagger$ \\
  & Recall@50 & 0.1147 & \underline{0.1918} & 0.1657 & 0.1905 & 0.1686 & 0.1416 &0.1511&	0.1588&0.1895 &\textbf{0.2042}$^\ddagger$\\
  \cmidrule(lr){2-12}
  & NDCG@10 & 0.0292 & \underline{0.0546} & 0.0433 & 0.0543 & 0.0437 & 0.0318 & 0.0373&	0.0413	&0.0536&\textbf{0.0583}$^\ddagger$ \\
  & NDCG@20 & 0.0375 & \underline{0.0684} & 0.0550 & 0.0675 & 0.0559 & 0.0422 & 0.0485&	0.0527	&0.0669&\textbf{0.0728}$^\ddagger$ \\
  & NDCG@50 & 0.0530 & \underline{0.0921} & 0.0772 & 0.0917 & 0.0780 & 0.0625 &0.0689&	0.0740&	0.0909 &\textbf{0.0982}$^\ddagger$ \\
  \midrule
  \multirow{6}*{Pinterest} & Recall@10 & 0.0771 & 0.0981 & 0.1012 & 0.1100 & 0.1140 & 0.0795 & 0.0813&	0.0976&	\underline{0.1143}&\textbf{0.1200}$^\ddagger$ \\
  & Recall@20 & 0.1278 & 0.1477 & 0.1625 & 0.1764 & 0.1692 & 0.1376 &0.1334&	0.1571	&\underline{0.1775} &\textbf{0.1776}  \\
  & Recall@50 & 0.2369 & 0.2414 & 0.2846 & \textbf{0.3063} & 0.2812 & 0.2658 & 0.2481&	0.2786	&\underline{0.3023} & 0.2911\\
  \cmidrule(lr){2-12}
  & NDCG@10 & 0.0486 & 0.0677 & 0.0647 & 0.0711 & \underline{0.0836} & 0.0483 & 0.0511&	0.0619	&0.0752 &\textbf{0.0853} \\
  & NDCG@20 & 0.0645 & 0.0834 & 0.0840 & 0.0921 & \underline{0.1010} & 0.0666 &0.0675&	0.0807	&0.0952 &\textbf{0.1035}$^\ddagger$ \\
  & NDCG@50 & 0.0916 & 0.1065 & 0.1144 & 0.1243 & \underline{0.1288} & 0.0983 &0.0959&	0.1107&	0.1261 &\textbf{0.1313} \\
  \bottomrule
\end{tabular}
\end{table*}

\end{document}